# Bulk-fragment and tube-like structures of Au$_N$ ($N$=2-26)

Wei Fa,[*] Chuanfu Luo, and Jinming Dong[†]

*Group of Computational Condensed Matter Physics,*

*National Laboratory of Solid State Microstructures,*

*and Department of Physics, Nanjing University, Nanjing, 210093, China*

(Dated: September 27, 2005)

## Abstract

Using the relativistic all-electron density-functional calculations on the Au$_N$ ($N$=2-26) in the generalized gradient approximation (GGA), combined with the guided simulated annealing (GSA), we have found that the 2- to 3-dimensional structural transition for Au$_N$ occurs between $N = 13$ and 15, and the Au$_N$ (16$\leq N \leq$25) prefer also the pyramid-based bulk fragment structures in addition to the Au$_{20}$. More importantly, the tube-like structures are found to be the most stable for Au$_{24}$ and Au$_{26}$, offering another powerful structure competitor with other isomers, e.g., amorphous, bulk-fragment, and gold-fullerene. The mechanism to cause these unusual Au$_N$ may be attributed to the stronger $s-d$ hybridization and $d-d$ interaction enhanced by the relativistic effects.

## I. INTRODUCTION

In recent years, gold clusters have attracted a surge of interests due to their fundamental importance and tremendous potential applications in novel nanoscale devices[1,2] and catalysts.[3] It is well-known that there exists a close relationship between their geometries and corresponding properties, but the existing experimental techniques cannot conclusively determine the structures of clusters. Thus, further numerical simulations on them become a useful method to study their geometrical structures by comparison of the calculated properties with the measured ones, such as those successfully done in alkali-metal clusters.[4]

So far, systematic numerical studies on $Au_N$ based on density-functional theory (DFT) have been carried out for $N \leq 20$,[5–11] which found strikingly that $Au_N$ prefer planar structures up to an unexpectedly large size of $N = 13$ than that expected for other metal clusters.[6] For 3-dimensional (3D) geometries, both the DFT calculations and the empirical potentials predicted that the amorphous packing is dominant for $Au_N$ with $N >14$,[5,12–14] in which the possible fragment structures of the face-centered cubic (fcc) lattice of bulk gold have never been taken into account. More recently, the photoelectron spectroscopy (PES) experiment revealed that $Au_{20}$ has an extremely large energy gap between the highest occupied molecular orbital (HOMO) and the lowest unoccupied molecular orbital (LUMO), and the further relativistic DFT calculations showed a new quasi-tetrahedral geometry for $Au_{20}$,[15] which is, in fact, a small piece of bulk gold with a slight structural relaxation, and more stable than the previously suggested "lowest-energy" compact $C_1$ or $C_{2v}$ structures.[5,13] The discovery of $T_d$ $Au_{20}$ challenged previous structural predications of $Au_N$ with 3D geometries. Therefore, it is natural to ask if there is a possibility to find other gold clusters with bulk-fragment structures.

In this paper, we have reconsidered the lowest-energy structures of $Au_N$ up to 26 atoms. It is found unexpectedly that the bulk fragments are stable for $Au_N$ ($N$=16-23, 25), and more importantly, the hollow tube-like structures for $Au_{24}$ and $Au_{26}$. Details of the calculations are described in Sec. II. In Sec. III, we discuss our results for the equilibrium geometries of $Au_N$ with ($N \leq$26) atoms. Section IV presents the concluding remarks.

## II. COMPUTATIONAL METHOD

We have made calculations on the atomic and electronic structures of $Au_N$ based upon the spin-polarized relativistic all-electron DFT in the generalized gradient approximation (GGA)[16] using the DMol$^3$ package.[17] A double-numerical basis set together with polarization functions was chosen to describe the electronic wave functions. Self-consistent field procedures were done with a convergence criterion of $10^{-6}$ a.u. on the total energy and electron density. Geometry optimizations were performed by the BFGS algorithm[18] without any symmetry constraints. The initial candidate geometries included not only a large number of available structures in the literature, but also new ones obtained by an unbiased global search with the guided simulated annealing (GSA),[19] which was made both in a 3D full space and on the fcc lattice points in order to obtain some bulk-fragment structures. The interaction between gold atoms in the Au$_N$ is represented by a tight-binding potential.[20] For each Au$_N$ with $N >5$, we chose at least eight lower-energy isomers with different symmetries from the GSA global search, and then picked up some planar structures to form a set of different structures, in which the global minimum of Au$_N$ was finally determined according to the optimization results of DMol$^3$. Finally, the frequency check has been performed to validate the stability of Au$_N$. No imaginary frequencies were obtained, indicating the obtained structures are true minima rather than saddle points on the potential energy surface.

### III. RESULTS

The second order total energy differences of Au$_N$ are presented in Fig. 1 as a function of their sizes, in which shown are also their ground-state structures.[21] Especially impressive is appearance of the pyramid-based bulk fragment and tube-like structures for the Au$_N$ with $N >15$. It has been found that the 2D → 3D transition is completed via close-flat structures from 13 to 15 atoms. On the other hand, it can be also seen from Fig. 1 that, except Au$_{18}$ and Au$_{22}$, the even-$N$ Au$_N$ are relatively more stable than the neighboring odd-$N$ ones.

The ground-state structures and energetic ordering of the smaller Au$_N$ ($N$ = 2-12) are very close to the existing DFT-GGA results.[6–8,22] For example, for the Au$_7$, the edge-capped rhombus with $C_s$ symmetry is 0.11 eV lower in energy than the edge-capped square with $C_{2v}$ symmetry, but for the Au$_7^-$, the latter is more stable than the former by 0.06 eV, showing a good agreement with Ref. 6. Our calculations corroborate that the surprising stability of the planar gold clusters continues up to 12 atoms, all of which are composed of a slice of the (111) face of bulk gold with a small structural relaxation.

Our DFT-GGA calculations found that the 3D structures of Au$_N$ set in the size range of $N=$ 13-15, preferring the close-flat configurations. $Au_{13}$ is composed of a distorted $Au_6$ and an edge-capped square planar $Au_7$. $Au_{14}$ can be viewed as a concave planar $Au_{11}$ closed by a gold triangle, from which $Au_{15}$ can be obtained by adding one more atom aside. There are also several kinds of competitive isomers of $Au_{13}$ and $Au_{15}$ for their lowest-energy configurations. For example, another $C_s$ three-capped cross-hexagonal planar structure of $Au_{13}$ is only 8 meV higher in energy than its ground state, and there exist also its many other isomers, forming a series of $Au_{13}$ with almost a continuous energy distribution.[23] It is found that although the 2D $\rightarrow$ 3D transition of $Au_N$ happens at $N=12$, the planar structures are still competitive for even larger sizes up to 15 atoms, making its energy difference between 2D and 3D structures to be only 0.13 eV (corresponding to a mean vibrational temperature of 77 K[24]). Also, it is worth noticing that a distorted bulk fcc type structure of $Au_{15}$ exists between the close-flat structure and the planar one, implying a structural transition towards the bulk fragment.

Since the most stable $Au_{20}$ is a tetrahedral pyramid, we naturally try to construct $Au_N$ ($N >$15) from the special pyramid-based structures. As expected, it is found by a careful global search that the most stable structures of $Au_{16}$-$Au_{23}$ and $Au_{25}$ can be obtained by removing or adding atoms to the $T_d$ $Au_{20}$, all of which are the fragment of bulk gold with a structural distortion. The ground-state geometries of $Au_{16}$–$Au_{19}$, with $T_d$, $C_{3v}$, $C_{2v}$, and $C_{3v}$ symmetries, respectively, come from the pyramid $Au_{20}$ by removing four, three, two, and one apex atom of $Au_{20}$. We should specially mention that the fragment structure of $Au_{16}$, with the same $T_d$ symmetry as $Au_{20}$, has a large HOMO-LUMO gap of 1.339 eV and is found to be strongly favored over all the other isomers, showing its high stability and chemical inert. For example, the space-filled compact structures proposed in Refs. 5, 12 are 0.86 and 0.94 eV higher in energy, respectively, than the $T_d$ $Au_{16}$. On the other hand, for the $Au_N$ with $N >$20, our calculations found the bulk fragment is still the ground state of $Au_{21}$-$Au_{23}$ and $Au_{25}$ with one, two, three, and five atoms, respectively, added in an additional (111) face next to the bottom layer of the pyramid $Au_{20}$.

In addition, compared with the amorphous and bulk-fragment ones, the distorted tube-like structures seem to become another stable structure competitor. For example, we found other two $Au_{22}$ isomers: one is amorphous as that found in Ref. 12, and the other is tube-like, produced by closing a (8, 4) gold nanotube (GNT),[25] among which the former is 0.64 eV higher, but the latter is only 0.06 eV higher in energy than the lowest-energy $Au_{22}$.

Therefore, with increasing gold atoms, probably the tube-like structures would become more and more competitive, even relative to the bulk fragment. As expected, both $Au_{24}$ and $Au_{26}$ are eventually found to prefer the hollow tube-like structures to the amorphous and pyramid–based ones. The ground-state $Au_{24}$ is obtained by a short segment of the (8, 0) GNT capped by two rhombi on both sides. To check stability of the tube-like $Au_{24}$, we have compared it with its two compact isomers with $C_s$ and $C_{3v}$ symmetry, respectively, predicted to be the "global" minima in previous calculations,[12,13] which are found to be 0.65 and 0.82 eV higher in energy than the tube-like structure, respectively, but energy-close to the pyramid-based isomer. Also, the HOMO-LUMO gap of the tube-like $Au_{24}$ equals 0.525 eV, which is larger than that of its compact isomers (about 0.2 eV). Similarly, the lowest-energy structure of $Au_{26}$ can be derived from closing a distorted piece of the (7, 0) GNT with two equilateral triangles. The stable tube-like $Au_{24}$ and $Au_{26}$ show another powerful competitor with other isomers, e.g., the amorphous, bulk-fragment, and fullerenic-like ones.[26,27]

To further clarify the geometrical features of $Au_N$, we have shown their average bond length and coordinate number in Fig. 2, from which it is clearly seen that the planar structures have the shortest bond lengths and the smallest bond number, leading to appearance of a distinct step at the 2D → 3D transition point of $N$=12.[28] On the other hand, by comparing the pyramid-based or tube-like structures of $Au_N$ with those in Ref. 12, we have found the former has shorter average bond length, showing the stronger relativistic contraction of the inter-atomic distances for $Au_N$. For example, the $T_d$ $Au_{16}$ has an average bond length of 2.72 Å, which is shorter than that with a $C_s$ symmetry of Ref. 12 (2.76 Å).

In order to identify clearly whether the relativistic effect (RE) plays also an important role in the new tube-like $Au_N$ we have found, we have calculated the electronic density of states (DOS) for the stable tube-like $Au_{24}$,[29] the same configurations of $Ag_{24}$, and non-relativistic $Au_{24}$ (denoted as NR-$Au_{24}$, re-optimized by the DFT including all-electron but NO RE), which are shown in Fig. 3. There exists a distinctly larger difference between the DOS of $Au_{24}$ valence band and those of the other two, especially near the Fermi level. By comparing Fig. 3(a) with 3(b) and 3(c), it is obvious that the RE plays two important roles in $Au_N$: 1) It causes a wider $d$-band and increases heavily the partial DOS of the $d$ orbital in the energy range of -3-0 eV, which comes from the stronger overlap between the $5d$ orbitals of neighboring gold atoms, produced by the shorter bond length. 2) It makes the $s$-band extend to lower energy regions, causing further an enhanced $s-d$ hybridization. All of the two effects are also found in the anionic Au$_7$.[11] Another good example to illustrate the influence of RE in gold clusters on their geometries is the DFT study on 20-atom coinage metal clusters.[30] By comparing the molecular orbitals of the Au$_{24}$, NR-Au$_{24}$, and Ag$_{24}$, shown in the right panel of Fig. 3, it is found that although their LUMOs are somewhat similar, there are big differences between their HOMO states, especially between the Au$_{24}$ and the other two, showing the larger $d$ electron contributions in the HOMO of Au$_{24}$. Finally, we would like to emphasize that because the tube-like structures have been found also for other non-relativistic elements, e.g., carbon, boron and nitrogen, etc, showing their rather universality, the origin of the tube-like Au$_N$ and GNTs could be more complex than the only RE. Therefore, further more experimental and theoretical studies may be needed to make it clear.

## IV. CONCLUSION

In summary, we have shown that the gold clusters prefer 3D structures, beginning with a close-flat configuration from Au$_{13}$, after which there exist more stable 3D distorted bulk fcc type structures in addition to the pyramid Au$_{20}$ for the Au$_N$ with $16 \leq N \leq 25$. Most importantly, a new kind of Au$_N$ with a tube-like structure has been found, in which Au$_{24}$ is the smallest one found until now. Such a tube-like structure may be more frequently found for the larger and larger Au$_N$. The ultimate driving force behind these unusual Au$_N$ ($N \leq 26$) structures may be the stronger RE in gold, causing a wider $d$-band and stronger $s-d$ hybridization. Our results might stimulate further experimental and theoretical studies to identify these non-amorphous or tube-like Au$_N$, and find their potential applications in physics, chemistry, and material sciences.

## ACKNOWLEDGMENTS

This work has been supported by the Natural Science Foundation of China under Grant No. A040108, and also by the state key program of China through Grant No. 2004CB619004. The DFT calculations were made on the SGI Origin-3800 and 2000 supercomputer.

---

[*] Electronic address E-mail:`wfa@nju.edu.cn`

[†] Corresponding author E-mail:`jdong@nju.edu.cn`

**Figure Captions**

FIG. 1. (Color online) Second order total energy differences of $Au_N$ vs cluster size. Also shown are the lowest-energy structures with even-atoms $Au_N$ drawn above the curve, while odd-atoms ones below the curve. To display the structures clearly, the top view is used for the smaller $Au_N$ with $N \leq 15$, and the side view for the bigger ones. The vertical dashed lines indicate the transition from planar to pyramid-like structures.

FIG. 2. (Color online) Average bond length and coordination number vs cluster size, represented by solid and open circles, respectively. For comparison, also shown are the average bond length values of the "lowest-energy" compact configurations found by Ref. 12 in the size range of 16 to 26, denoted by rhombus.

FIG. 3. (Color online) DOS of the tube-like clusters: (a) $Au_{24}$, (b) NR-$Au_{24}$, and (c) $Ag_{24}$. The Fermi level is shifted to zero, denoted by the vertical dashed line. The partial DOS of the $s$ orbital is shown by the shaded areas. The right panel gives the corresponding HOMO and LUMO orbitals of the three clusters. The isodensity surfaces correspond to 0.025 $e/au^3$.

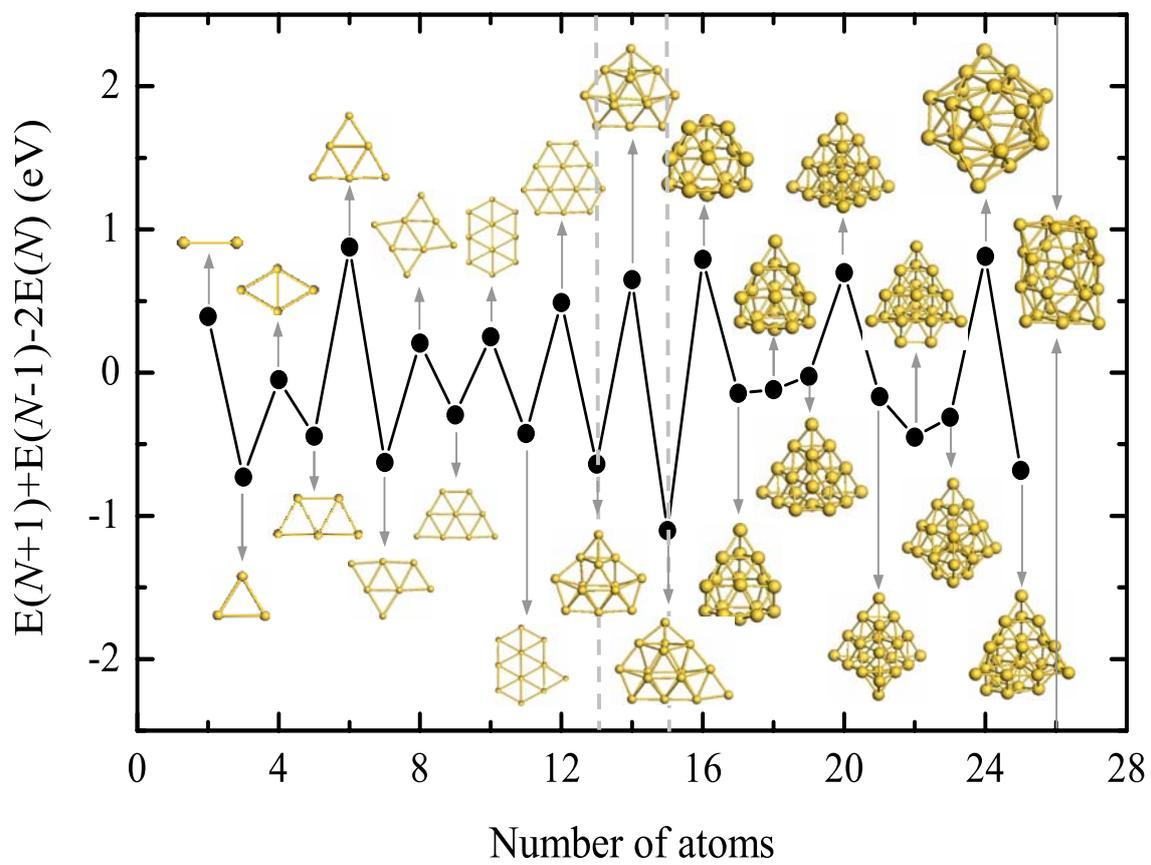

Figure 1

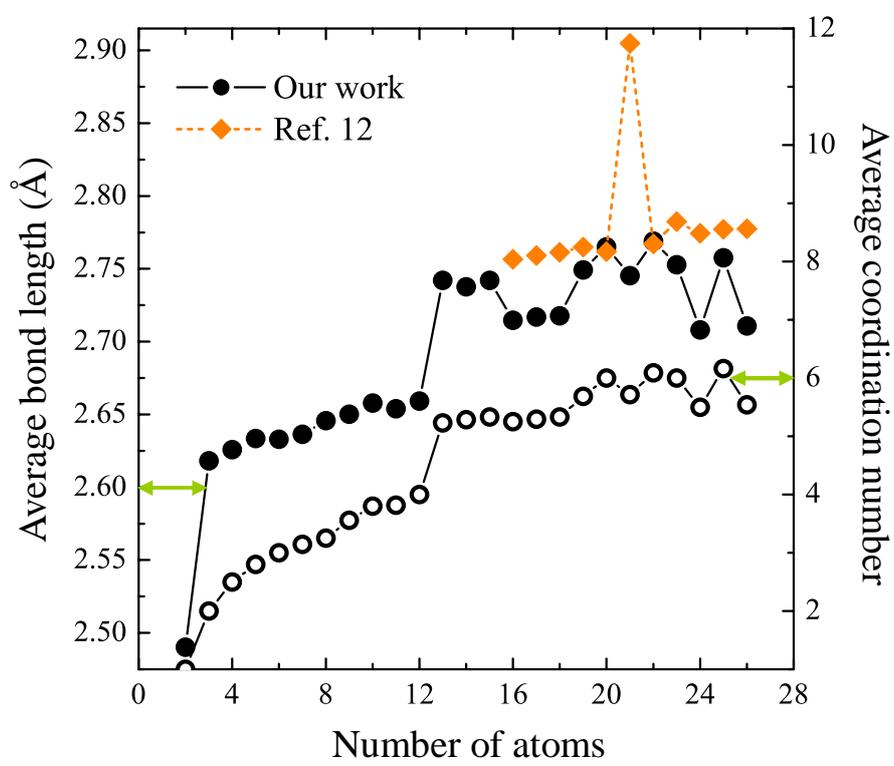

Figure 2

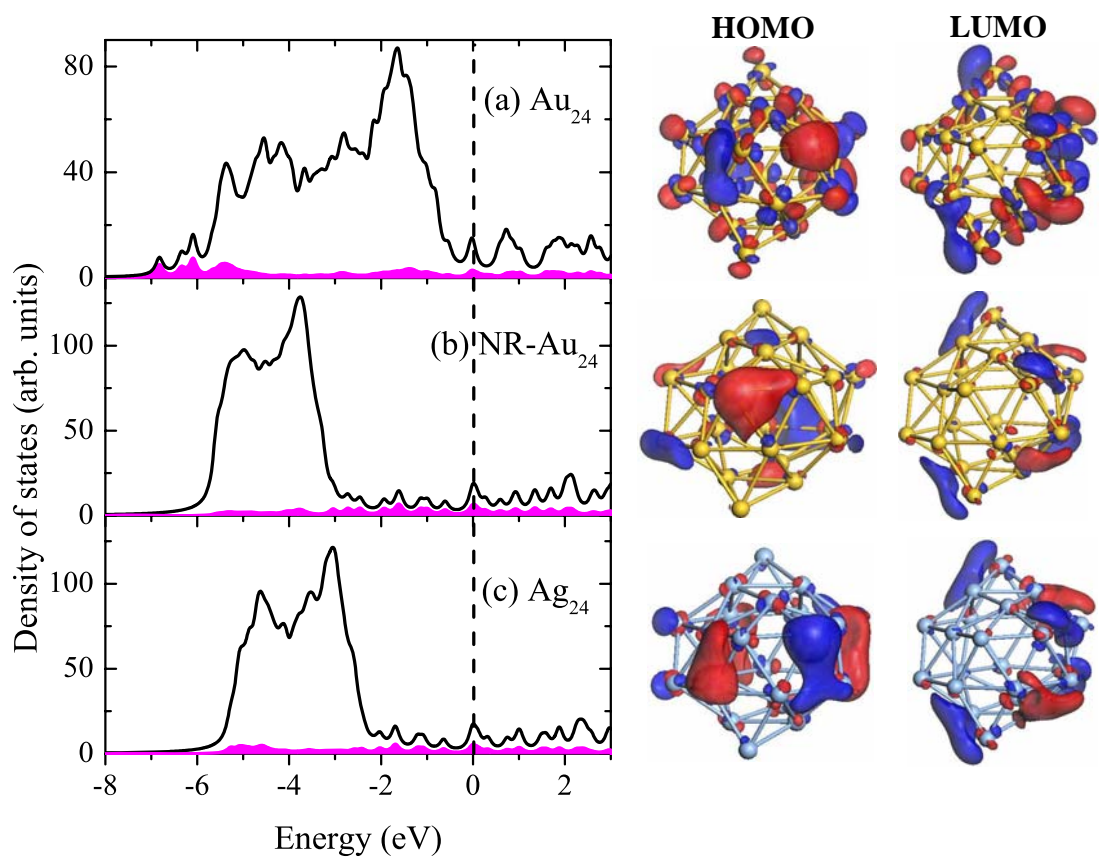

Figure 3